\documentclass[aps,prl,twocolumn,superscriptaddress,groupedaddress]{revtex4}  
\usepackage{graphicx}  
\usepackage{dcolumn}   
\usepackage{bm}        
\usepackage{amssymb}   
\usepackage{amsmath}   

\begin{document}


\title{Temporal Simultons in Optical Parametric Oscillators}
\affiliation{Edward L. Ginzton Laboratory, Stanford University, Stanford, CA, 94305}
\affiliation{Physics Department, ETH Zurich, 8093 Zurich, Switzerland}

\author{Marc Jankowski} \email[e-mail: ]{marcjank@stanford.edu}\affiliation{Edward L. Ginzton Laboratory, Stanford University, Stanford, CA, 94305}
\author{Alireza Marandi} \email[e-mail: ]{marandi@stanford.edu}\affiliation{Edward L. Ginzton Laboratory, Stanford University, Stanford, CA, 94305}
\author{C.R. Phillips} \affiliation{Physics Department, ETH Zurich, 8093 Zurich, Switzerland}
\author{Ryan Hamerly} \affiliation{Edward L. Ginzton Laboratory, Stanford University, Stanford, CA, 94305}
\author{K.A. Ingold} \affiliation{Edward L. Ginzton Laboratory, Stanford University, Stanford, CA, 94305}
\author{R.L. Byer} \affiliation{Edward L. Ginzton Laboratory, Stanford University, Stanford, CA, 94305}
\author{M.M. Fejer} \affiliation{Edward L. Ginzton Laboratory, Stanford University, Stanford, CA, 94305}       
\date{\today}

\begin{abstract}
We identify and demonstrate a regime of operation in optical parametric oscillators (OPOs) in which the formation of temporal simultons produces stable half-harmonic pulses. Simultons are simultanous bright-dark solitons of a signal field at frequency $\omega$ and the pump field at $2\omega$ which form in a quadratic nonlinear medium. The formation of simultons in an OPO is evidenced by sech$^2$ spectra with broad instantaneous bandwidths which increase with pump power, and large slope efficiencies. In contrast to conventional synchronously pumped OPOs, operation in this regime is achieved by using relatively large parametric gains, and a low finesse resonator detuned to a slightly longer roundtrip time than the pump repetition period. In the experiment we achieve sub-50 fs sech$^2$ pulses with a slope efficiency of 570\%, and a conversion efficiency of 55\% as a result of formation of simultons in an OPO. We verify the distinct features of OPO operation in the simulton regime analytically and numerically. These results represent a fundamental shift in the understanding and design of OPOs, in which the nonlinear dynamics that arise during asynchronous operation can be used to efficiently generate few-cycle long-wavelength frequency combs.
\end{abstract}

\maketitle

\textit{Introduction.---}The formation and propagation of stable femtosecond optical pulses in nonlinear resonators has played a key role in the development of frequency comb sources. Modern techniques for generating femtosecond optical pulses use nonlinear dynamics, including the formation of dissipative Kerr solitons and similaritons in mode-locked lasers and Kerr microresonators\cite{Haus,Nail-DispSolitons,KerrSolitons,SimilaritonLaser}. These pulse formation mechanisms occur due to an interplay between gain, loss, dispersion, and $\chi^{(3)}$ nonlinearities, and have been successfully used to generate few-cycle pulses and phase-stabilized frequency combs in the range from 400 nm-3.5 $\mu$m\cite{KerrSolitons,Cherenkov,MIRcomb,CrZnSe}. While considerable effort is being invested to extend frequency combs to other wavelength ranges\cite{MIRcombs}, such operation requires overcoming the challenges associated with developing either new broadband laser gain media or high finesse resonators.

Optical parametric oscillators (OPOs) based on the $\chi^{(2)}$ nonlinearity offer a compelling source of frequency combs across infrared wavelengths. The broad bandwidth tunable gain available from optical parametric amplification (OPA) can generate wavelengths where broadband laser gain media are not readily available, and the large parametric gains available in $\chi^{(2)}$ systems allow for oscillation to occur without high finesse cavities. While pulse formation mechanisms are well studied in continuous-wave-pumped degenerate OPOs\cite{Fabre,Longhi,Wabnitz,Wabnitz2}, in which a $\chi^{(2)}$ resonator pumped at $2\omega$ generates a resonant half-harmonic at $\omega$, to date these systems have not yet achieved mode-locked femtosecond pulses by using such dynamics\cite{Ulvila1,Ulvila2}. Synchronously pumped degenerate OPOs have been used successfully to generate half-harmonic combs, but their pulse formation mechanisms are less understood. Key results include the demonstration of instantaneous octave-spanning-spectra\cite{Leindecker}, few-cycle pulses\cite{Sorokin}, intrinsic phase and frequency locking\cite{Coh} which translates the coherence properties of the pump source onto the half-harmonic signal, and conversion efficiencies as high as 65\%\cite{Cascade}. 
Recent work suggests a number of competing pulse formation mechanisms exist in such OPOs\cite{Ryan}.

In this letter we identify a regime of operation in a near-synchronously pumped degenerate OPO in which stable half-harmonic pulses are formed by temporal simultons. This letter will proceed in three parts. (i) We develop a reduced model of simulton formation, and explain the characteristics of simultons in the context of OPO operation. (ii) We present experimental results, and identify signatures of simulton formation. (iii) Numerical simulations are used to better understand the underlying dynamics and are shown to capture the behavior exhibited by the OPO. Based on this agreement, we are able to connect the proposed intracavity simulton dynamics to the observed behavior of the OPO.

\begin{figure}
\centering\includegraphics[width = 8.6cm]{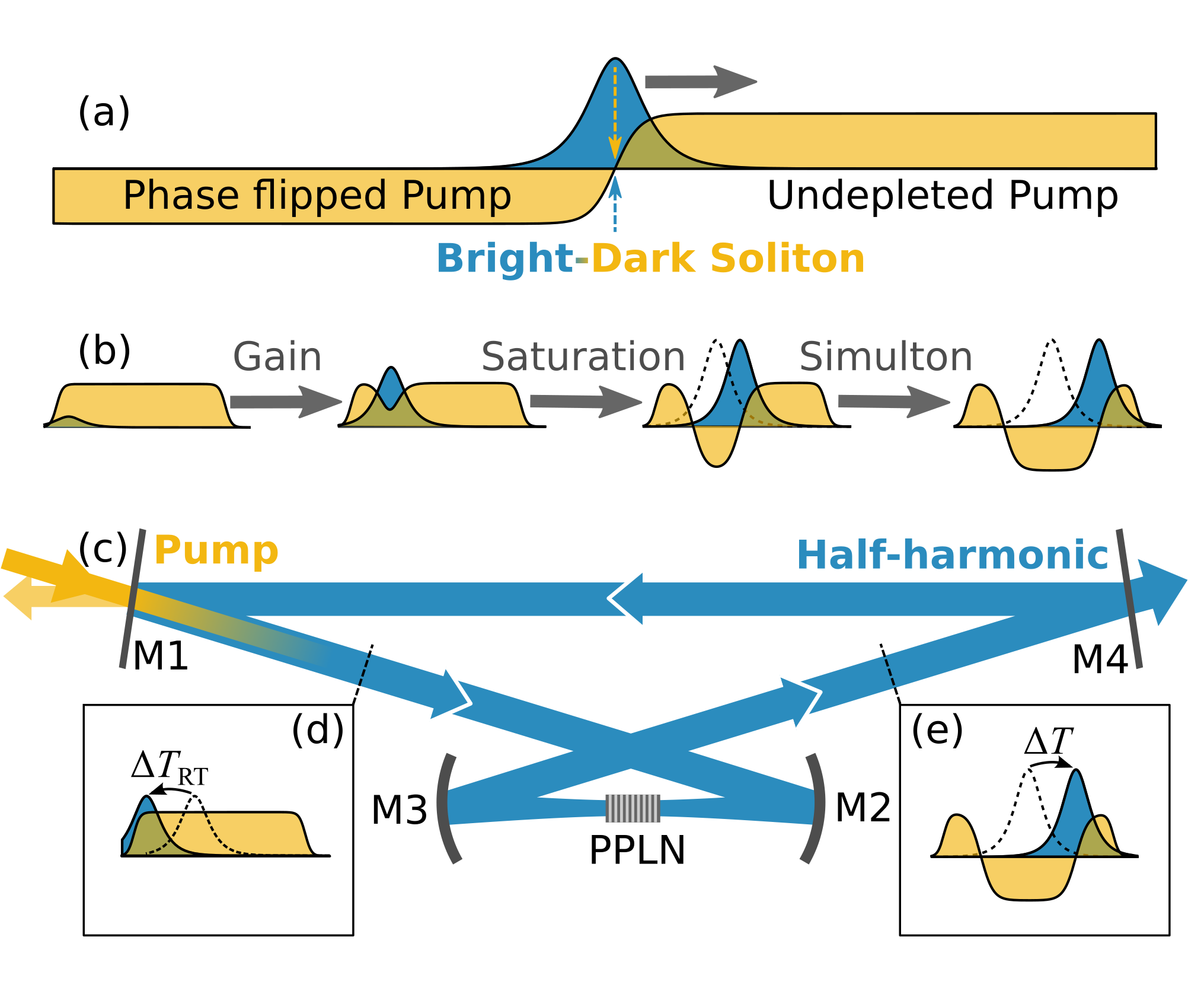} 
\caption{ (a) Field envelopes of a simulton, showing the sech-pulse signal (blue) and tanh pump (orange). (b) Evolution of pump and signal fields from a linear OPA into a simulton. Dotted line - the evolution of the signal field undergoing linear temporal walkoff. (c) Schematic of the synchronously pumped OPO cavity. Cavity length adjustments are made by mounting M1 on a piezo stage. (d-e) Motion of intracavity signal relative to a perfectly synchronous half-harmonic pulse undergoing linear propagation (dotted lines) (d) After M1 the signal acquires a small delay, $\Delta T_\mathrm{RT}$, relative to the incoupled pump due to the timing mismatch. (e) After optical parametric amplification the signal acquires a nonlinear group delay shift $\Delta T$ due to simulton formation, which compensates the timing mismatch $\Delta T_\mathrm{RT}$.}
\end{figure}

\textit{Theory.---} Temporal simultons are simultaneous bright-dark solitons of the signal at $\omega$ and the pump at $2\omega$, which occur in a degenerate OPA due to group velocity mismatch and gain saturation\cite{Akh,Trillo}. The coupled wave equations are
\begin{subequations}
\begin{align}
\partial_z A_\omega(z,t) &= \kappa A_{2\omega}A_\omega^*,\label{cwe_1}\\
\partial_z A_{2\omega}(z,t) &= -\Delta\beta' \partial_t A_{2\omega}-\kappa A_\omega^2,\label{cwe_2}
\end{align}
\end{subequations}
where we have shifted the time coordinate to be co-moving with the group velocity of the signal wave, and include a $\pi/2$ phase in the pump envelope to make the equations of motion and their solutions real, assuming perfect phasematching for OPA.  $A_\omega$ is the field envelope, normalized such that $|A_\omega|^2$ is the instantaneous power of the $\omega$ wave, $\kappa$ is the nonlinear coupling, and $\Delta\beta'=v_{g,2\omega}^{-1}-v_{g,\omega}^{-1}$ is the group velocity mismatch. The simulton solution is given by \cite{Akh}
\begin{subequations}
\begin{align}
A_\omega(z,t) &= \frac{a}{\sqrt{2\tau}}\mathrm{sech}\left(\frac{t-T}{\tau}\right),\label{sim_1}\\
A_{2\omega}(z,t) &= -E_{2\omega}\tanh\left(\frac{t-T}{\tau}\right),\label{sim_2}
\end{align}
\end{subequations}
where $a^2=2\gamma(\Delta\beta'+\gamma \tau)/\kappa^2$ is the signal pulse energy, $\gamma =\kappa E_{2\omega}$ is the small-signal gain coefficient, $\tau$ is the pulse width, and $T=-\gamma \tau z$ represents a shift in the signal pulse relative to linear propagation due to gain saturation. Simultons occur when the leading edge of a bright sech$^2$ signal pulse depletes a quasi-continuous wave pump, and the trailing edge converts back to the pump frequency through second harmonic generation (SHG) with a $\pi$ phase relative to the undepleted pump (Fig.1(a)). The pump forms a tanh$^2$ dark soliton coupled to the bright sech$^2$ signal pulse, and the pair co-propagate with an intensity dependant velocity which exceeds that of either wave, $v_{g,\mathrm{sim}}^{-1}=v_{g,\omega}^{-1}-\gamma \tau$.


We generalize this solution to include gain and loss using the manifold projection method described in \cite{Ryan}, and obtain the evolution of the parameters $a$, $\tau$, and $T$ of the sech-like signal pulse from Eq. \ref{sim_1}. When $\Delta\beta'z \gg \tau$ and $\Delta\beta' \gg \gamma\tau$, $a(z)$, $T(z)$, and $\tau(z)$ evolve as
\begin{subequations}
\begin{align}
\partial_z a &= \gamma a\left[1-\frac{a^2}{a_\mathrm{sim}^2}\right],\label{dzA}\\
\partial_z T &= -\gamma \tau \frac{a^2}{a_\mathrm{sim}^2},\label{dzT}\\
\partial_z \tau &= 0.\label{dztau}
\end{align}
\end{subequations}
Here $a_\mathrm{sim}^2 = 2\Delta\beta'\gamma /\kappa^2$ is the simulton energy, and we have approximated the pump as a flat-top pulse, $E_{2\omega} = \max(A_{2\omega}(0,t))$. Eqs. (\ref{dzA}-\ref{dztau}) can be understood in two limits. When $a\ll a_\mathrm{sim}$ we recover the evolution of a degenerate OPA with an undepleted pump. The signal is amplified as $a(z)=a(0)e^{\gamma z}$, and propagates with a linear group velocity, i.e. $\partial_z T=0$. When $a=a_\mathrm{sim}$ we recover the simulton solution (Eqs. \ref{sim_1}-\ref{sim_2}) with $\Delta\beta' \gg \gamma\tau$. In the limit of the approximations made here, the simulton solution is a stable attractor. If a sech signal pulse is seeded into a degenerate OPA such that $a>a_\mathrm{sim}$, it will transfer energy to the pump through SHG until a simulton is formed. Eqs. (\ref{dzA}-\ref{dzT}) can be solved for the full evolution of a dissipative simulton, resulting in:
\begin{subequations}
\begin{align}
a(z) &= \frac{a(0)e^{\gamma z}}{\sqrt{1+\frac{a^2(0)}{a_\mathrm{sim}^2}\left(e^{2\gamma z}-1\right)}},\label{gain}\\
\Delta T(z) &= \tau\mathrm{ln}\left(\frac{a(0)e^{\gamma z}}{a(z)}\right).\label{advance}
\end{align}
\end{subequations}
$\Delta T(z)= T(z) - T(0)$ is the shift in group delay accumulated due to nonlinear acceleration in a single pass through the OPA crystal. A schematic of simulton formation is illustrated in Fig.1(b), showing the evolution of the pump and signal from an undepleted OPA into a simulton. The signal is seen to undergo linear temporal walkoff due to group velocity mismatch and extract gain until the pump is depleted. Once depleted, the pump forms a co-propagating dark soliton, and the pair accelerate to the simulton velocity.

Fig.1(c-e) show the dynamics of a simulton OPO. On each round trip, a new pump pulse enters the cavity through the input coupler, M1 (Fig.1(c)), and the signal accumulates a small group delay $\Delta T_\mathrm{RT}$, hereafter referred to as the timing mismatch, due to an offset between the pump repetition period and the cold-cavity round trip time (Fig.1(d)). After passing through the OPA crystal, labelled PPLN, the signal is amplified, and accumulates a simulton group advance $\Delta T$(Fig.1(e)). The signal is partially out-coupled through M4, with a fraction $R$ of the power returning to M1. Simulton formation in an OPO is a double balance of energy and timing in which the gain extracted over an OPA crystal of length $L$ balances the cavity loss, $a^2(0) = Ra^2(L)$, and the simulton acceleration balances the timing mismatch, $\Delta T(L)=\Delta T_{RT}$. When the timing condition is satisfied, the signal becomes synchronous with the pump and forms a half-harmonic pulse which inherits both it's carrier-envelope offset frequency and comb spacing from the pump. The equations for steady state, with Eqs. (\ref{gain}-\ref{advance}) determine the simulton pulsewidth:
\begin{subequations}
\begin{align}
\tau &= \frac{2\Delta T_\mathrm{RT}}{2\gamma L+\mathrm{ln}(R)}.\label{tau}
\end{align}
\end{subequations}

The pulsewidth of a simulton OPO is seen to shrink with increasing pump power, in contrast to the conventional $\tau\propto\sqrt{P}$ ``box-pulse'' scaling developed in \cite{Ryan}. For positive detunings ($\Delta T_\mathrm{RT}>0$), the simulton group advance allows for the formation of half-harmonic pulses which are synchronous with the pump at multiple cavity lengths. Negatively detuned ($\Delta T_\mathrm{RT} < 0$) simultons cannot form when $\Delta\beta'>0$ since pump depletion only provides a group advance for the signal pulse. Instead, the OPO operates in a non-degenerate regime analyzed in \cite{Ryan}. The timing mismatch is thus a critical design parameter which determines both the mode of operation and the bandwidth of the OPO.

\begin{figure}[t]
\centering\includegraphics[width=8.6cm]{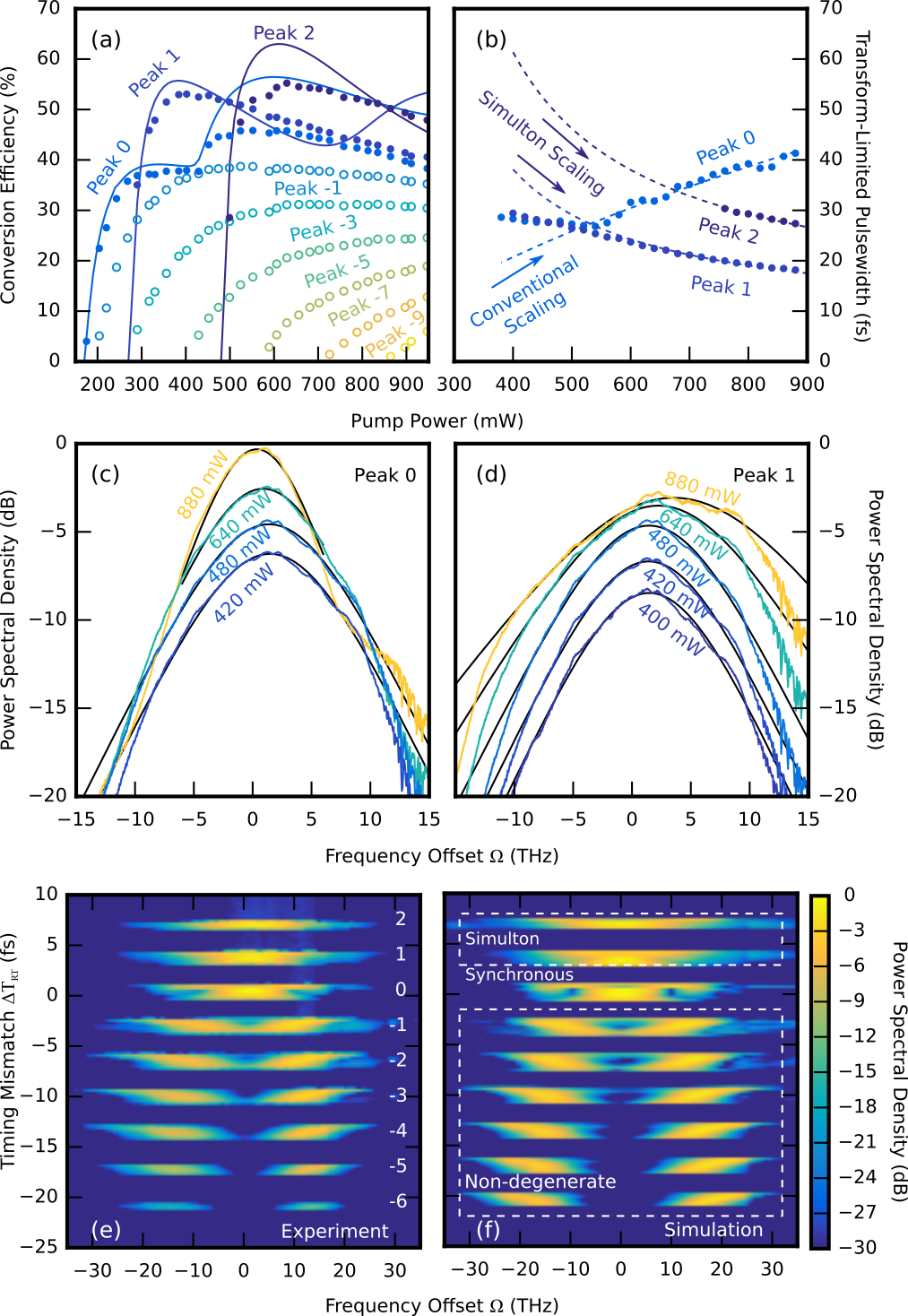} 
{\tiny \caption{ (a) Measured conversion efficiency for each resonance, with the resonances enumerated relative to perfect synchronization. Positive ``peak'' numbers correspond to a long cavity. Solid lines represent numerical simulations of peak 0, 1, and 2, and empty circles denote nondegenerate operation. (b) The scaling with pump power of the transform limited pulsewidth (3 dB) for peaks 0-2. For large powers, peak 0 shows an increase in pulsewidth in accordance with \cite{Ryan}, while peaks 1 and 2 show a monotonic decrease in pulsewidth which agrees well with Eq. \ref{tau}. (c-d) (Color lines) Spectra recorded as a function of pump power for peak 0 and 1 respectively. Each curve is labelled with the corresponding pump power used in the experiment. (Black lines) Sech$^2$ fits to the experimental spectra. (e) Measured signal spectrum as a function of timing mismatch for a 550 mW pump, with the associated peak numbers. (f) Simulated signal spectrum in dB relative to the maximum value, as a function of timing mismatch, with the three regimes we have identified indicated.}}
\end{figure}

\textit{Experimental Results.---}We study the behavior of an OPO as the timing mismatch is varied around perfect synchronization with the pump. The OPO cavity consists of a the same bowtie resonator as \cite{Cascade} with a tunable round-trip delay of $\sim$4~ns (Fig.1(c)), a large output coupling of $(1-R)=65\%$ for the signal, and a $1/e^2$ beam radius of 10~$\mu$m for the pump. OPA occurs in a 1-mm-long Brewster-cut MgO-doped periodically poled lithium niobate (PPLN) crystal placed at the focus between M2 and M3. The PPLN crystal has a poling period of 31.8~$\mu$m to phasematch degenerate OPA of a signal at 2090~nm, and is pumped by 70~fs pulses at 250~MHz produced by 1045-nm mode-locked Yb-fiber laser (Menlo Systems Orange A) with an average power of up to 950~mW.  The OPO oscillates around cavity lengths where the signal acquires a phase shift of 0 or $\pi$ relative to the pump on each round trip, leading to a discrete set of resonances whose behavior depends strongly on the timing mismatch.

We first consider the resonance at which the OPO cavity is most nearly synchronized to the pump repetition rate, labelled Peak 0 in Fig.2(a). The synchronous peak has the lowest threshold, 175~mW, a slope efficiency of $158\%$, and a peak conversion efficiency of $46\%$. Moreover, it exhibits a sech$^2$ spectrum, which loses bandwidth with increasing power (Fig.2(b-c)), in accordance with the conventional box-pulse scaling\cite{Ryan}. As the cavity is positively detuned ($\Delta T_\mathrm{RT}>0$), two more resonances are found, labelled Peak 1 and 2 in Fig.2(a). These ``long cavity'' resonances have irregularly spaced thresholds as $\Delta T_\mathrm{RT}$ becomes increasingly positive, and have measured slope efficiencies as high as $570\%$, with peak efficiencies of $55\%$. Peaks 1 and 2 exhibit sech$^2$ spectra which monotonically increase in bandwidth as the pump power is increased (Fig.2(b,d)) in accordance with the simulton scaling, Eq. \ref{tau}. The spectra deviate from the exponential tails of a sech$^2$ spectrum beyond $\pm$10 THz due to atmospheric absorption around 1850 nm. Peak 1 achieves a 3dB bandwidth as high as 240~nm, which can support pulses as short as 19~fs. When the cavity is negatively detuned ($\Delta T_\mathrm{RT}<0$) the OPO transitions to non-degenerate operation and the spectra split into a distiguishable signal and idler (Fig.2(e)). The peaks in the non-degenerate regime exhibit thresholds which increase uniformly as $\Delta T_\mathrm{RT}$ becomes increasingly negative, slope efficiencies less than $40\%$, and conversion efficiencies less than $40\%$ (Fig.2(a)). We therefore identify three regimes of operation associated with the timing mismatch: synchronous ($\Delta T_\mathrm{RT}=0$), non-degenerate ($\Delta T_\mathrm{RT}<0$), and simulton ($\Delta T_\mathrm{RT}>0$).

\textit{Simulation.---} To better understand the dynamics which determine the three regimes of operation observed in the experiment and verify that the positively detuned resonances correspond to simulton operation, we study the OPO using numerical methods. The OPO is modeled as an OPA followed by a linear feedback loop. On each round trip, we solve Eqs. (\ref{cwe_1}-\ref{cwe_2}) using split-step Fourier methods including all dispersion orders. We model the feedback loop as a linear filter for the signal:
$$A_\omega^{(n+1)}(0,t) = \mathcal{F}^{-1}\lbrace \sqrt{R(\Omega)}e^{-i\phi(\Omega)}\mathcal{F}\lbrace A_\omega^{(n)}(L,t)\rbrace\rbrace.$$
The phase $\phi(\Omega)$ is measured relative to a half-harmonic signal which is perfectly synchronous with the pump
$$\phi(\Omega)=\phi_0 + \pi l + \Delta T_{\mathrm{RT}}\Omega + \Delta\phi(\Omega),$$
where $\phi_0$ represents an offset between the cavity resonances and the cavity length which synchronizes the pump and signal, $\Delta\phi(\Omega)$ represents the quadratic and higher order dispersion of the cavity mirrors, and $l=c\Delta T_\mathrm{RT} /\lambda_{2\omega}$ parameterizes the peak number as the cavity length is varied from perfect synchronization, with resonances centered on cavity lengths such that $l\in\mathbb{Z}$.

The solid lines in Fig.2(a) show the simulated conversion efficiency of the resonances in the synchronous and simulton regimes, and are shown to be in good agreement with the experimental thresholds and slope efficiencies. Deviations which occur at higher powers are likely due to radial variations in pump depletion not included in the simulaton. A simulation of the spectrum as a function of timing mismatch with parameters corresponding to the experiment is shown in Fig. 2(f), with the three regimes of operation indicated by the dashed boxes. The simulations show excellent agreement with the experimental data (Fig. 2(e)) in all three operating regimes. In the case of $\Delta T_\mathrm{RT}>0$, stable femtosecond half-harmonic pulses are generated through the formation of simultons.

\begin{figure}[t]
\centering\includegraphics[width=8.6cm]{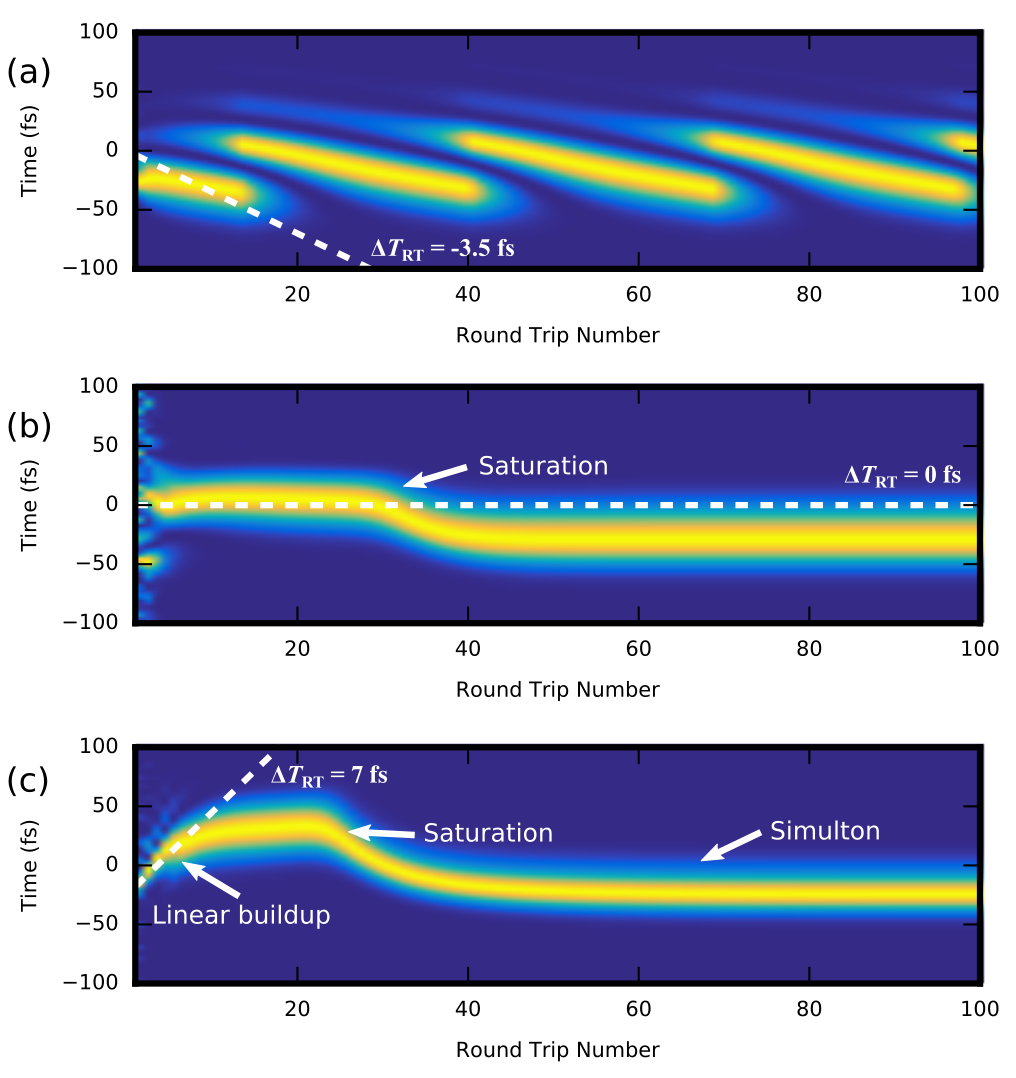} 
{\tiny \caption{ (a) Simulated evolution of the intracavity pulse intensity over many round trips for $\Delta T_\mathrm{RT}=$-3.5~fs showing the interference of a distinct signal and idler. The dotted line denotes the trajectory of linearly propagating half-harmonic signal pulse. (b) Pulse evolution for $\Delta T_\mathrm{RT}=0$~fs showing the formation of a half-harmonic pulse synchronized to the pump repetiton rate. (c) Simulated pulse evolution for $\Delta T_\mathrm{RT}=$7~fs showing the formation of a half-harmonic pulse which, upon depleting the pump, is able to accelerate forward in time and synchronize to the pump repetition rate.}}
\end{figure}

Having shown agreement between the numerical model and experiment, we now use the model to better understand the femtosecond pulse formation dynamics in the OPO. The evolution of the signal pulse is shown in Fig.3(a-c) for each of the three regimes of operation. Each round trip is recorded at the output of the OPA and normalized to its peak amplitude to visualize the pulse motion. The dashed white lines show the expected trajectory of a linearly propagating half-harmonic signal pulse, which acquires a delay $\Delta T_\mathrm{RT}$ on every round trip. In these figures, the time coordinates have been shifted such that a signal peaked at $t=0$ corresponds to a pulse walking symmetrically from the tail of the pump at $\Delta\beta'L/2$ to the leading edge of the pump $-\Delta\beta'L/2$. For a negatively detuned peak (Fig.2(a)), the pulse envelope shows a 10~THz modulation in time (vertical fringes), resulting from interference of a signal and idler split from degeneracy by $\pm$10~THz. The interference fringes are seen to shift on each round trip, corresponding to a $\pm$10~MHz offset of the signal and idler carrier-envelope-offset frequency $f_\mathrm{ceo}$ from that of a half-harmonic signal (horizontal fringes). When the cavity length is tuned into synchronization (Fig.2(b)), the signal builds up without any motion relative to the pump until the pump saturates, shifting the signal forward in time until a new steady state is found on the leading edge of the pump. The evolution for a positively detuned cavity is shown in Fig.2(c). The signal initially tracks the trajectory of a linearly propagating pulse, shifting towards the tail of the pump. Once the signal is intense enough to deplete the pump, it accelerates, becoming faster than would be possible under linear propagation, and thereby reaches a steady state, synchronous with the pump repetition rate.

The surprising behavior exhibited by the long cavity resonances, namely a nonlinear acceleration of the signal pulses, verifies that the OPO dynamics in this regime correspond to simulton formation. Furthermore, the full numerical model facilitates an intuitive picture of the behavior of the simulton peaks. The large thresholds and slope efficiencies of the simulton peaks are due to the pulsewidth of the pump. When $a\ll a_\mathrm{sim}$ the signal pulse will accumulate many successive group delays due to the timing mismatch, and experience a decrease in gain due to a reduction in the temporal overlap with the pump. Since simulton operation requires the signal to be bright enough to deplete the pump, threshold then corresponds to the condition that the signal builds up from quantum noise to the simulton energy before the gain seen by the signal pulse becomes less than the cavity loss. Once this condition is satisfied, the signal accelerates back into the pump and depletes it, leading to large slope efficiencies.

\textit{Conclusion.---}We have demonstrated the formation of simultons in a near-synchronously pumped OPO when the round trip delay of the cavity is longer than the repetition period of the pump. Simultons are observed to  generate stable phase-locked sech$^2$ pulses with large instantaneous bandwidths, suggesting them as a promising source of mode-locked femtosecond pulses and frequency combs at previously inaccessible wavelengths. The scaling of pulsewidth with pump power in Eq. \ref{tau} suggests that simultons might be used to realize single-cycle pulses, and the numerical methods detailed above confirm that in principle sub 2-cycle pulses are possible with conversion efficiencies $>80\%$. Design rules for the cavity dispersion and finesse based on the reduced model discussed here will be the subject of future publications.


The authors would like to acknowledge support from DARPA DODOS award  2014003913-03, NSF award ECCS-1609688, and NSF-BSF award PHY-1535711.   

\end{document}